\documentclass[twocolumn]{aastex62}
\usepackage{latexsym}
\usepackage{color}
\usepackage{amsmath}
\usepackage{amssymb}
\usepackage{graphicx}
\usepackage{aas_macros}

\newcommand{\Ms} {$\rm{M_{\odot}}~$}

\usepackage{hyperref}

\submitjournal{ApJ}
\shortauthors{Schauer et al.}
\shorttitle{Earendel - a Pop III star?}

\begin{document}

\title[Earendel --- a Pop III star?]
{On the probability of the extremely lensed $z$=6.2 Earendel source being a Population~III star} 

\correspondingauthor{Anna T. P. Schauer}
\email{anna.schauer@utexas.edu}

\author[0000-0002-2220-8086]{Anna T. P. Schauer}
\affiliation{Department of Astronomy, 
University of Texas at Austin,
TX 78712, USA}

\author[0000-0003-0212-2979]{Volker Bromm}
\affiliation{Department of Astronomy,
University of Texas at Austin,
TX 78712, USA} 

\author[0000-0002-7339-3170]{Niv Drory}
\affiliation{McDonald Observatory, 
University of Texas at Austin,
TX 78712, USA}

\author[0000-0002-9604-343X]{Michael Boylan-Kolchin}
\affiliation{Department of Astronomy,
University of Texas at Austin,
TX 78712, USA} 

\begin{abstract}
The recent discovery of the extremely lensed Earendel object at $z=6.2$ is remarkable in that it is likely a single star or stellar multiple, observed within the first billion years of cosmic history. Depending on its mass, which is still uncertain but will soon be more tightly constrained with the James Webb Space Telescope, the Earendel star might even be a member of the first generation of stars, the so-called Population~III (Pop~III). By combining results from detailed cosmological simulations of the assembly of the first galaxies, including the enrichment of the pristine gas with heavy chemical elements, with assumptions on key stellar parameters, we quantify the probability that Earendel has indeed a Pop~III origin.  We find that this probability is non-negligible throughout the mass range inferred for Earendel, specifically ranging from a few percent at the lower-mass end to near unity for some Pop~III initial mass function (IMF) models towards the high-mass end of the allowed range. For models that extend the metal-enriched IMF to $500$\,M$_\odot$, the likelihood of Earendel being a Pop~III star stays at the few to ten percent level. We discuss the implications of such a discovery for the overall endeavor to probe the hitherto so elusive first stars in the universe.
\end{abstract}

\section{Introduction}\label{introduction}
The recent discovery of a star-like object at redshift $z=6.2$, termed Earendel (Old English for `morning star'), which is observable due to extreme lensing by a massive intervening galaxy cluster has been a surprise to the astronomical community \citep{welch22}. This object, inferred to be an individual star or a small stellar multiple, was found by the Hubble Space Telescope (HST) in the Reionization Lensing Cluster Survey (RELICS; \citealt{coe19}). With the available data, the exact nature of the star remains uncertain, but it is constrained to be massive, in excess of $\sim 50 {\rm \,M}_{\odot}$. A particularly intriguing possibility is that this star could be a member of the hitherto elusive Population~III (Pop~III), formed out of the primordial hydrogen-helium gas, before any enrichment with heavy chemical elements \citep[][]{bond1981}. Upcoming observations with the James Webb Space Telescope (JWST) will measure the star's metallicity, thus deciding its precise physical nature. In this Letter, we will discuss the probability that Earendel is indeed a Pop~III star. 
 
 To see why this would be such a surprising result, let us recall the standard model of first star formation within a $\Lambda$CDM universe 
 \citep[e.g.][]{fg11,volkerreview13,haemmerle20}. In broad terms, this model predicts that the first (Pop~III) stars form in minihalos at $z\sim20-30$, with an initial mass function (IMF) that is top-heavy. Such massive stars have very short lifetimes, and would quickly die in energetic supernova explosions. The concomitant metal enrichment would rapidly establish a local metallicity floor, such that the next generation of star formation would already give rise to metal-enriched Pop~I/II stars \citep[e.g.][]{pallottini2014,jaacks19,magg22}. Because of this prompt transition in star formation mode, even ultra-deep observations with the JWST are not expected to detect Pop~III stars. To reach the very beginning of cosmic star formation, a future `ultimately-large' telescope (ULT) would thus be required \citep{Angel+2008,ult}, for example a 100m-diameter liquid-mirror design on the lunar surface. 
 
 The endeavor to extend the high-redshift frontier has been one of the key motivations of observational cosmology, with the ultimate goal of reaching back to Pop~III star formation \citep[][]{barkana2001}. Traditionally, the prime targets were luminous quasars, powered by accretion onto supermassive black holes (SMBHs), with a current record of $z\sim 7.6$ for J0313-1806 \citep{wang21}. More recently, galaxies have overtaken quasars in marking the redshift frontier, due to the increasing scarcity of SMBHs at early cosmic times \citep{woods19}. Of particular note are the extreme H-band drop-out candidates HD1 and HD2, with photometrically estimated redshifts of $z\sim 13$ \citep{harikane22}. It is an open question whether these sources are powered by intense starbursts, a central SMBH, or a combination thereof \citep{pacucci22}. For the starburst interpretation, a top-heavy IMF may be required, which in turn may point to a Pop~III origin. 
 
 Alternatively, the high-redshift universe may be probed with rare, but hyper-energetic transient events, linked to the deaths of individual Pop~III stars \citep{lazar22}, such as pair-instability supernovae (PISNe) and gamma-ray bursts (GRBs). The search for high-$z$ sources will benefit from any flux amplification through gravitational lensing. It had previously been hypothesized that in extreme cases of magnifications of order a few $\sim$1,000s, compact star clusters, including possibly Pop~III ones, may be accessible in future wide-field surveys \citep[e.g.][]{zackrisson15}.  
 
 Complementary to the high-redshift strategy of reaching Pop~III, we may search for more recent, `fossil' survivors. One possibility is that Pop~III stars continue to form at lower redshifts, in rare pockets of primordial gas, which could then be identified as PISN explosions \citep[e.g.][]{sc05,tss09,liu20}. Even if such metal-free pockets cannot persist at $z\lesssim 6$, we may still be able to discover Pop~III fossils, if the primordial IMF extended to low masses, $\lesssim 0.8\, M_{\odot}$, with corresponding stellar lifetimes in excess of the current age of the universe \citep[e.g.][]{frebel2010,hb15}.
 
 In the following, we will combine results from cosmological simulations that constrain the properties of the first galaxies, including the inferred host galaxy for the Earendel star, with a consideration of its key stellar properties, to assess how likely it is Pop~III. In light of the standard model expectation and the corresponding pathways to discovery, such a `shortcut' to the pristine beginning of star formation would be truly remarkable. 

\section{Assessing a Pop~III origin}\label{prob}
While \cite{welch22} mention the possibility of Earendel being a Pop~III star, 
they did not provide a quantitive probability of this scenario. In what follows, we estimate the probability that Earendel is indeed a Pop~III star.
Our calculation is based on three components: the stellar IMF, the incidence of Pop~III star forming regions in cosmological host halos, and the lifetime over which the stars can be observed. 
\subsection{Host halo environment}
After the formation of the first stars, the universe becomes metal enriched by heavy elements, primarily through supernova explosions \citep[reviewed in][]{karlsson2013}. Some metal-free pockets exist down to lower redshift \citep{tss09}. The corresponding probability of encountering metal-free star forming regions in a given host galaxy as a function of halo mass and redshift has been quantified by \cite{liu20}. Their study is based on cosmological simulations of high-$z$ star formation, including detailed modeling of radiative feedback, metal enrichment through supernova winds, and the fine-grained mixing of heavy elements. 

Earendel likely is part of the Sunrise Arc galaxy with a stellar mass of $3\times10^7$\,M$_\odot$ \citep[][]{welch22}, corresponding to a lower limit for the halo mass (baryons and dark matter) of $\sim 10^9$\,M$_\odot$, when assuming an extremely large star formation efficiency of 20\% and a baryon fraction of 16\% \citep{Planck2016}. The inferred halo mass therefore always exceeds the cosmological filtering mass for fully-ionized gas, placing the Sunrise Arc in the ``post-reionization'' range considered in \cite{liu20}. From their figure~15a, we find a Pop~III occupation fraction of $f_\mathrm{\rm PopIII} \approx 0.008$ for this case, providing an estimate for the likelihood that Earendel's host galaxy exhibits metal-free star formation. 
\subsection{Initial mass function}
The shape of the IMF describes the probability that a star has a certain mass. While the IMF of metal-enriched stars has been directly observed, we need to rely on indirect modeling and simulations for the shape and mass range of the Pop~III IMF. Specifically, for Pop~II (or Pop~I) stars, we adopt the commonly used lower-mass limit of 0.1\,M$_\odot$. For the shape, we employ a Larson-type expression: 
\begin{equation}
\frac{\mathrm{d}N}{\mathrm{d}M}\propto M^{-2.35} \times \exp{[-(m_{\rm char}/M)]}\mbox{\ ,}
\label{equ:larson}
\end{equation}
with a characteristic mass of $m_{\rm char}=0.35$~\Ms \citep[e.g][]{chabrier2003}. For Pop~III stars, we follow the recent study of \cite{rossi21}, who inferred  a lower mass  limit of 1\,\Ms from the absence of detected metal-free stars in dwarf galaxies (see also \citealt{hartwig15a} for a similar study regarding the Milky Way). This is in agreement with numerical simulations, which further describe the Pop~III IMF as top-heavy. 
We adopt two different Pop~III IMFs: a Larson-type one (see Equ.~\ref{equ:larson}) with a characteristic mass of $m_{\rm char}=10$~\Ms, and 
a log-normal IMF (as employed by e.g. \citealt{magg16}). 
\begin{equation}
\frac{\mathrm{d}N}{\mathrm{d}\log M}\propto {\rm const}.
\label{equ:lognormal}
\end{equation}
These choices of Pop~III IMFs are commonly used in the literature, and here serve as limiting cases, with a relatively steep Pop~II like high-mass slope for the Larson case and a very top-heavy IMF in the log-normal case. 

For the upper mass limit, we formally adopt 500\,M$_\odot$ for both the Pop~II and Pop~III case. For Pop~II, such unrealistically high upper limit affects the normalization only marginally, and we keep the high-mass end at the same value for both stellar populations for simplicity. 
However, we note the the IMF of Pop II and Pop I stars usually does not exceed 150\,M$_\odot$ \citep[][]{figer2005}, and the highest mass star observed so far has a mass of 250$\pm$120\,M$_\odot$ in the luminous star-forming region W49 \citep{wu16}. In addition to the first model, where we simply extend the IMF to 500\,M$_\odot$, we also include a Pop~II IMF with an exponential cutoff at 150\,M$_\odot$, reflecting empirical evidence for such a mass limit for metal-enriched stars \citep[e.g.][]{weidner2004}. We approximately model this by multiplying the Pop~II IMF by $\exp[1-(M/150 \mathrm{\,M}_\odot)^2]$ for masses exceeding 150\,M$_\odot$. 
Finally, we normalize the IMFs by integrating $M(\mathrm{d}N/\mathrm{d}M) \mathrm{d}M$ over the mass range of $[0.1, 500]$\,M$_\odot$ and $[1, 500]$\,M$_\odot$ for Pop~II and Pop~III, respectively. 

\begin{figure}
    \centering
    \includegraphics[width=0.5\textwidth]{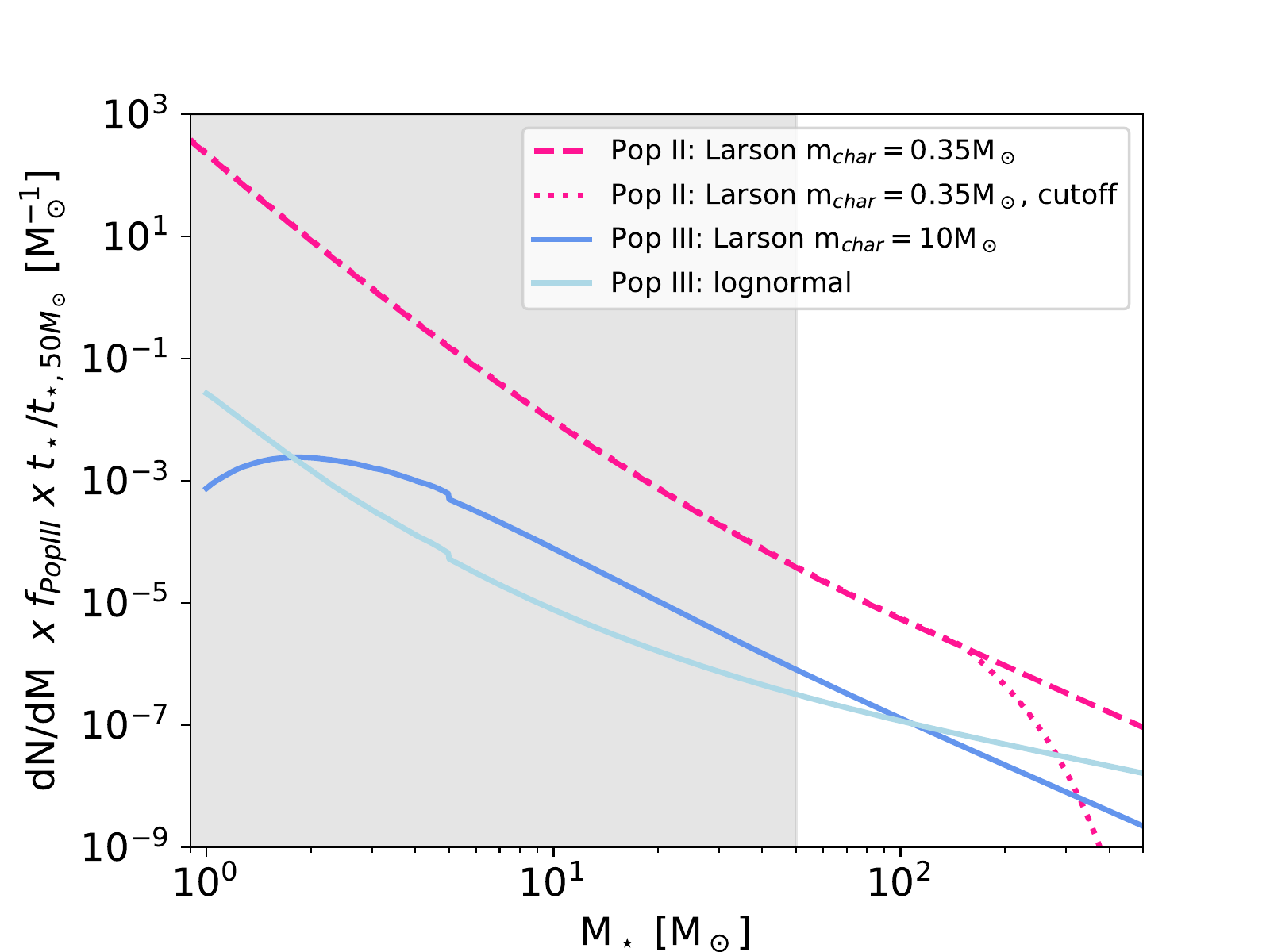}
    \caption{Weighted number density of stars in the Earendel host galaxy vs. stellar mass. {\it Blue and light blue lines:} Pop~III case for the Larson and log-normal IMF. {\it Dotted and dashed pink lines:} Pop~II case with and without the empirical high-mass cutoff beyond 150\,M$_\odot$. The white region indicates the allowed mass range for Earendel between 50--500\,M$_\odot$.}
    \label{fig:nm}
\end{figure}

\subsection{Stellar lifetimes}
We furthermore need to consider stellar lifetimes to estimate how likely it is to observe a star of a given mass. As the lifetime of a star decreases with mass, we are more likely to observe a low-mass star than a high-mass star. We use Pop~III stellar evolutionary models from \cite{schae02} and \cite{marigo01} for stars with masses above and below 5\,M$_\odot$, respectively. Lifetimes of Pop~II stars are approximated for a 1/50~Z$_\odot$ stellar population from \cite{schae02}, keeping in mind that metal-enrichment generally increases the stellar lifetime by a few 10\%. As the fit provided by \cite{schae02} for Pop~II stars is only valid up to 150\,M$_\odot$, we use the Pop~III model multiplied by $t_{\rm II}(150\,M_\odot) / t_{\rm III}(150\,M_\odot)$ for M$_{\rm II} > 150$\,M$_\odot$. 
\subsection{Results}
Finally, we investigate the probability that a star (cluster) observed at redshift $z=6.2$ is a Pop~II or Pop~III star. In Figure~\ref{fig:nm}, we show the weighted number density of these stars as a function of mass for our three cases, $\mathrm{d}N/\mathrm{d}M\times f_{\rm PopIII} \times t_\star/t_{\star, 50M_\odot} $ as a function of stellar mass. The Pop~II number density stays above the Pop~III one for both Pop~III models, except at the highest masses where the exponential cutoff dominates. 

\begin{figure}
    \centering
    \includegraphics[width=0.5\textwidth]{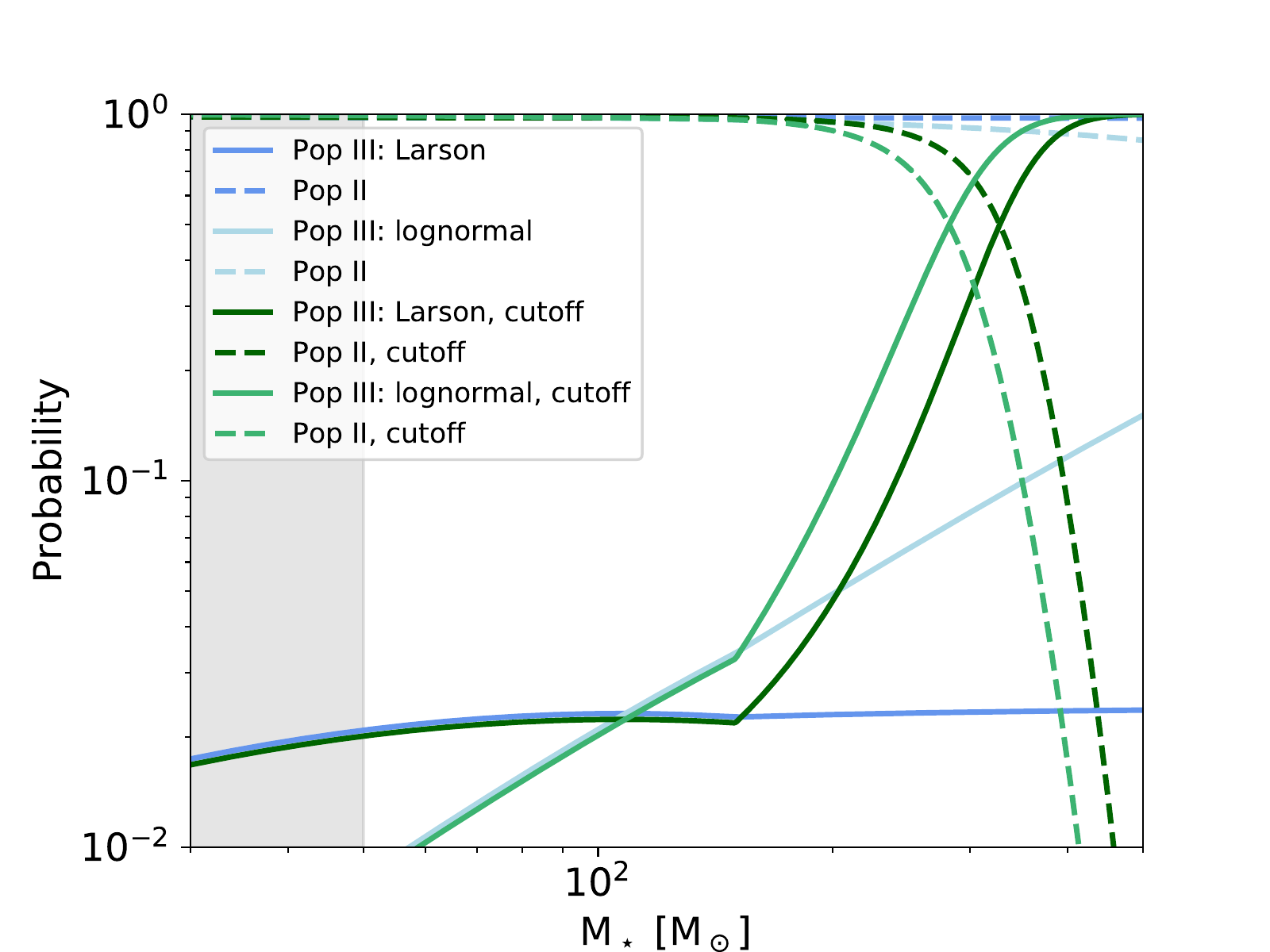}
    \caption{Probability of Earendel being a Pop~II star ({\it dashed lines}) or a Pop~III star ({\it solid lines}) as a function of stellar mass. The two Pop~III cases are a Larson-type IMF with a characteristic mass of $m_\mathrm{char} =0.35$\,M$_\odot$ ({\it blue}) and a log-normal IMF ({\it light blue}). While the Larson-IMF model stays nearly constant at a Pop~III probability of 1-2\%, the log-normal IMF case rises to 15\%. If we include an exponential cutoff at the high-mass end of the Pop~II IMF, the probability for a Pop~III star exceeds the Pop~II probability at 280\,M$_\odot$ (log-normal IMF, {\it green}) or 330\,M$_\odot$ (Larson IMF, {\it dark green}), respectively. 
    }
    \label{fig:prob}
\end{figure}

In Figure \ref{fig:prob}, we show the probability that a star detected in the Sunrise Arc galaxy is a Pop~III or Pop~II star, for both cases of the adopted Pop~III IMF. The probability is calculated by the ratios of the Pop~II and Pop~III weighted number densities shown in Figure \ref{fig:nm}, assuming that Earendel is either a single star or a star cluster dominated by one massive star. Without a high-mass cutoff imposed for Pop~II stars, the probability for Earendel being a Pop~II star is always larger than 85\%, even for the highest stellar mass of 500\,M$_\odot$. If the Pop~III IMF follows a Larson-type expression with characteristic mass $m_{\rm char} = 10$\,M$_\odot$, the Pop~III probability stays relatively constant between 1\% and 2\%. If the Pop~III IMF follows a log-normal distribution, Earendel is more likely to be a Pop~III star. Beyond 60\,M$_\odot$, the probability for a Pop~III star exceeds 1\% and reaches 15\% at the highest mass-limit for Earendel.

As stated above, the existing evidence indicates that metal-enriched star
formation does not extend to extremely high masses. Modeling this semi-empirical constraint for Pop~II with an exponential cutoff for high-mass stars leads to a strong increase in the Pop~III probability beyond 150\,M$_\odot$, exceeding the Pop~II probability at 280\,M$_\odot$ for a log-normal IMF, or 330\,M$_\odot$ for a Larson IMF, respectively. 
This result remains valid for a wide range of Pop~III IMFs in Larson-type form with slopes varying from 2.35 to 1 and characteristic masses down to 1\,M$_\odot$. 
The mass above which a Pop~III origin for Earendel becomes more probable than a Pop~II one is of order $\approx$ 300\,M$_\odot$.
Even for a slope of 2.35 and a characteristic mass of 1\,M$_\odot$, Earendel is more likely a Pop~III star if it exceeds $\sim$350\,M$_\odot$.
Evidently, for the highest masses, Earendel's probability of being a Pop~III star is close to unity. This conclusion relies on the uncertain theoretical predictions for the upper mass limit of Pop~III stars \citep[e.g.][]{omukai2003}. The underlying physics of increased accretion rates and reduced radiation pressure in primordial gas, however, does favor Pop~III in reaching such large stellar masses. Overall, we conclude that there is a non-negligible chance that Earendel is indeed a Pop~III star, although the most likely outcome is still a Pop~II origin. This is quite surprising, given that Pop~III star formation is vastly outnumbered in terms of stellar mass fraction by metal-enriched populations, close to the epoch of reionization.

\section{Outlook}\label{outlook}

There is a significant chance that Earendel is a Pop~III star if it is indeed a single star of high mass. Given that the magnification is quite uncertain, determining the location of Earendel in the Hertzsprung-Russell diagram is only possible using spectroscopy. NIRSpec on board JWST has the required sensitivity to detect, e.g., mass-sensitive NLTE He~II emission features in Earendel's spectrum \citep{bkl01,nakajima22}, expected to have equivalent widths of $\gtrsim 100$\,\AA. If, on the other hand, Earendel is a low-number multiple, the most massive member star will dominate and there will remain some uncertainty, although it is still likely that a Pop~III origin can be confirmed or ruled out spectroscopically.

The discovery of a Pop~III star would be remarkable. 
This probability ranges from 1 to 100\%, depending on the mass inferred for Earendel and the actual Pop~III IMF, and is thus non-negligible. 
The detection of a high-redshift metal-free star would confirm our basic theory of cosmological structure formation and metal enrichment in an aging and expanding universe. More specifically, it would confirm the patchiness of early metal enrichment, as Earendel is embedded in a larger galaxy, which is likely metal-enriched. 
If Earendel were a Pop~III star, this would be the first constraint on the masses of metal-free stars. While many theoretical and computational studies predict a top-heavy IMF \citep{bromm01b, hir14}, we have no direct observational evidence yet to support this prediction. If, on the other hand, Earendel were a massive Pop~II star, we might have observed the most massive metal-enriched star in the universe. 

Even though lensing is a powerful tool in looking deeper into cosmic history, the accessible observable volume is small and we do not expect to sample a large number of Pop~III stars (\citealt{rydberg13}, see however another possible Pop~III cluster detection by \citealt{vanzella20}). 
To achieve a statistically meaningful survey, an even larger next-generation telescope, such as the ``Ultimately Large Telescope'' \citep{ult,Angel+2008}, is necessary to probe the properties of the first stars in the universe.

\section*{Acknowledgments}
We would like to thank the anonymous referee for constructive comments. 
AS\&MBK were partially supported by NSF CAREER award AST-1752913 and NASA grant NNX17AG29G. MBK also acknowledges support from NSF grants AST-1910346 and AST-2108962, and HST-AR-15809, HST-GO-15658, HST-GO-15901, HST-GO-15902, HST-AR-16159, and HST-GO-16226 from the Space Telescope Science Institute, which is operated by AURA, Inc., under NASA contract NAS5-26555.
VB acknowledges the Texas Advanced Computing Center (TACC) for providing HPC resources under XSEDE allocation TG-AST120024. 
\setlength{\bibhang}{2.0em}
\setlength\labelwidth{0.0em}

\label{lastpage}

\end{document}